# The glass transition and crystallization kinetic studies on BaNaB$_9$O$_{15}$ glasses

Rahul Vaish and K. B. R. Varma*

Materials Research Centre, Indian Institute of Science, Bangalore, India.

*Corresponding Author; E-Mail : kbrvarma@mrc.iisc.ernet.in;

FAX: 91-80-23600683; Tel. No: 91-80-22932914

**Abstract**

Transparent glasses of $BaNaB_9O_{15}$ (BNBO) were fabricated via the conventional melt-quenching technique. The amorphous and the glassy nature of the as-quenched samples were respectively, confirmed by X-ray powder diffraction (XRD) and differential scanning calorimetry (DSC). The glass transition and crystallization parameters were evaluated under non-isothermal conditions using DSC. The correlation between the heating rate dependent glass transition and the crystallization temperatures was discussed and deduced the Kauzmann temperature for BNBO glass-plates and powdered samples. The values of the Kauzmann temperature for the plates and powdered samples were 776 K and 768 K, respectively. Approximation-free method was used to evaluate the crystallization kinetic parameters for the BNBO glass samples. The effect of the sample thickness on the crystallization kinetics of BNBO glasses was also investigated.

# I. Introduction

Glass-ceramics are in general composites (glasses and crystallites) obtained by the controlled crystallization of the glassy materials. The glasses subjected to a carefully regulated heat-treatment schedule, result in the nucleation and growth of crystalline phases within the glass. Glass-ceramics have received considerable attention in the past few decades because of their promising applications in the fields of non-linear optics and pyroelectrics [1-3]. There are many advantages associated with glass-ceramics such as low/zero porosity and consequently high dielectric break down voltages, easy controllability of the properties and could be produced in a variety of sizes and shapes. Glassy dielectric materials comprising nano/micro crystallites of polar phases have been known to exhibit interesting physical properties which include piezoelectric, pyroelectric, electrooptic and non-linear optical [1,4-9]. In the last two decades, researchers have investigated glass-ceramic systems containing ferroelectric crystalline phases such as; $LiNbO_3$ [8], $SrBi_2Nb_2O_9$ [4] $Bi_2WO_6$ [5] etc. In the area of photonics, glasses comprising non-linear optical (NLO) crystals have received much attention, because these materials have high potential for laser hosts, tunable waveguides, tunable fiber gratings etc.

Borate-based compounds have attracted the attention of many researches because of their wide transmission window, moderate melting point, high chemical and mechanical stability, which make them useful for various applications [10]. Borate-based non-linear optical crystals, which include $Li_2B_4O_7$ [11], $La_2CaB_{10}O_{19}$ [12] and $CsLiB_6O_{10}$ [13] etc., were used to obtain UV light. We have been making systematic attempts to explore the possibilities of employing borate based glass systems comprising nano/micro polar crystals of the same phase for pyroelectric, electrooptic and non-linear optic device applications [1,4,5]. Recently, it was reported that BNBO crystallizes in noncentrosymmetric space group, R3c [14]. Therefore, we thought that it was worth attempting to grow nano/micro crystals of BNBO in their own glass matrix and visualize their non-linear optical and polar properties. To begin with one needs to have apropri knowledge about crystallization processes and mechanisms inorder to fabricate glass-ceramics of desired microstructure vis-à-vis transparency. We have fabricated surface crystallized BNBO glasses using ultrasonic treatment (UST) and the effect of the UST on crystallization behavior of the glasses was demonstrated using DSC [15]. In order to further our understanding of these glasses, the relation between the glass transition and the crystallization and the crystallization kinetics of the bulk and powdered samples are investigated in the present study. Differential Thermal Analysis (DTA) and Differential Scanning Calorimetry (DSC) could frequently be used to study the crystallization kinetics of the glassy materials [16-21]. For determining the kinetic parameters such as activation energy of crystallization ($E_c$) and $n$ in Johnson-Mehl-Avrami (JMA) equation [22-24], for the present BNBO glasses, non-isothermal approximation-free method is employed. Also the method of Chen [25] is used to estimate the above parameters. The activation energy associated with the glass-transition is determined using Kissinger [26] and Moynihan [27] methods. Heating rate dependent glass transition and crystallization temperatures are rationalized and correlated using Lasocka equation [28]. The Kauzmann temperatures [29], for the present glass system are calculated for the glass-plate and powdered samples using Lasocka equation. The details of which are reported in this paper.

**II. Experimental**

$BaNaB_9O_{15}$ (BNBO) glasses were fabricated via the conventional melt-quenching technique. For this, $BaCO_3$ (99.95%, Aldrich), $Na_2CO_3$ (99.9%, Merck) and $H_3BO_3$ (99.9%, Merck) were mixed and melted in a platinum crucible at 1373 K for 30 min. Melts were quenched by pouring on a stainless steel plate that was maintained at 423 K and pressed with another plate to obtain **0.5-2.5 mm** thick glass-plates. The DSC non-isothermal experiments were carried out using the power-compensated DSC (Model: Diamond DSC, Perkin Elmer). For glass transition analysis, the sample was heated and cooled from 738 K to 848 K at different heating rates (5, 10, 15 and 20 K/min). For crystallization studies, the plates of various thicknesses (0.5, 1.5 and 2.5 mm) and powdered (5-10 µm) glass samples were heated from 673 K to 973 K at the rate of 5, 10, 15 and 20 K/min. All the experiments were conducted in dry nitrogen ambience. The as-quenched glass samples (there were no nuclei present before DSC runs) weighing 20 mg were used for each experiments.

**III. Results and discussion**

The X-ray powder diffraction pattern that is obtained for the as-quenched sample confirms its amorphous nature (Fig. 1).

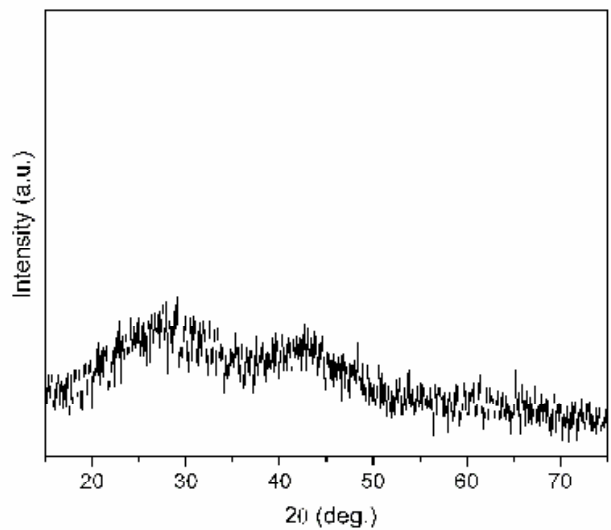

Fig. 1: XRD pattern for the as-quenched pulverized BNBO glass.

In Fig. 2, we show the typical DSC traces obtained for BNBO glass-plates of different thicknesses and powdered samples at a heating rate of 10 K/min. The onsets of the glass transition ($T_g$) and the crystallization temperature ($T_{cr}$) are identified as the temperatures corresponding to the intersections of the two linear portions of the transition (glass transition and crystallization) elbows of the DSC traces. As can be seen in Fig. 2, the glass transition temperature ($T_g$) is almost the same for all the samples (plates and powdered) whereas the crystallization occurs at a lower temperature for the powdered sample. **The crystallization** temperature is independent (within the limitations of the experimental errors) of the thickness of the glass plates under study. However, it is noticed that the width of the exothermic peaks decreases with increase in the thickness of the glass plates. It suggests that crystallization kinetic parameters of BNBO glasses are dependent on the thickness of the samples. The glass transition ($T_g$), the onset of crystallization temperature ($T_{cr}$), and crystallization peak width at half maximum ($?T_{FWHM}$) at a heating rate of 10 K/min for all the samples (glass-plates and powder) are summarized in table 1.

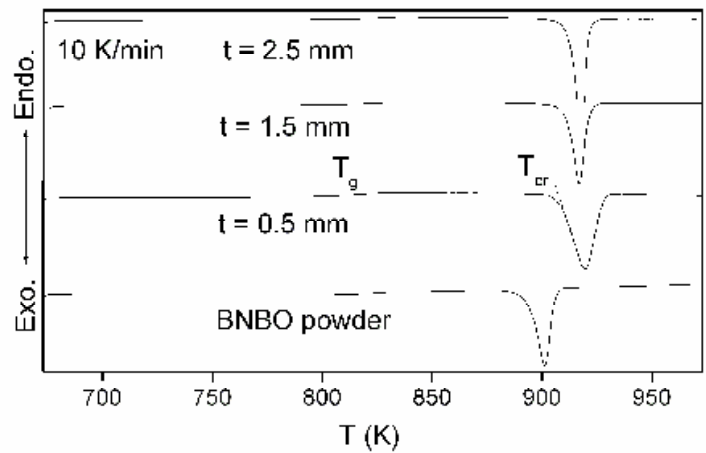

Fig. 2: DSC traces for the as-quenched samples of BNBO glass powder and plates of different thickness.

1. **Glass transition**

Glass transition studies are important from the view point of understanding the mechanism of glass transformations and in evaluating the structural rigidity of the glasses. The glass transition temperature reflects the strength or rigidity of the glasses. Fig. 3 shows the glass transition ranges that are obtained for BNBO glass-plates at different heating rates (5, 10, 15 and 20 K/min). The glass transition peak ($T_{gp}$) shifts to higher temperatures with increasing heating rate as shown in Fig. 3, indicating the kinetic nature of the glass transition.

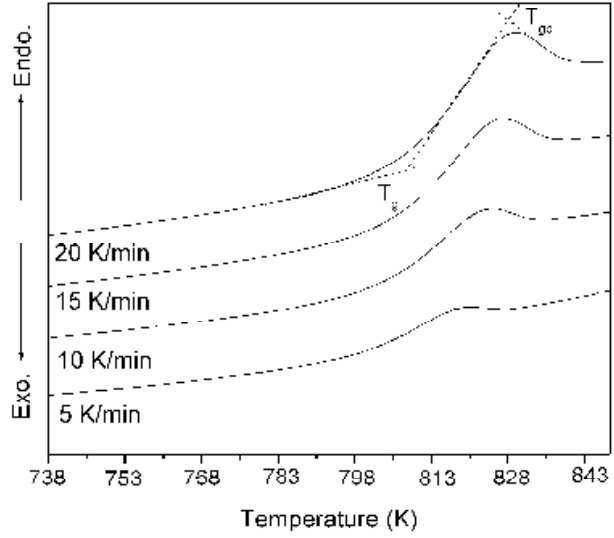

Fig. 3: Glass transition peaks for BNBO glass-plates at various heating rates.

The $T_g$ dependence on the heating rate ($a$) has been analyzed using the following three different approaches. The empirical relation between $T_g$ and $a$ according to Lasocka is

$$T_g = A_g + B_g \log \alpha \tag{1}$$

The value of $A_g$ is equal to the glass transition temperature for the heating rate (a) of 1 K/min and $B_g$ is constant for a given glass composition. The plot of $T_g$ versus log a for BNBO glasses and a theoretical fit (solid line) are shown in Fig. 4. The values that are obtained for $A_g$ and $B_g$ are 795 K and 9.9, respectively, for BNBO glasses. Therefore, the above relation can be written as

$$T_g = 795 + 9.9 \log \alpha \tag{2}$$

Eq. 2 is an excellent description of the dependence of $T_g$ on heating rate for BNBO glasses.

The second approach for analyzing $T_g$ or $T_{gp}$ is based on the Kissinger's formula, in which the $T_{gp}$ has a linear dependence on the heating rate, according to which

$$\ln\left(\frac{\alpha}{T_{gp}^2}\right) = const. - \frac{E_g}{RT_{gp}} \quad (3)$$

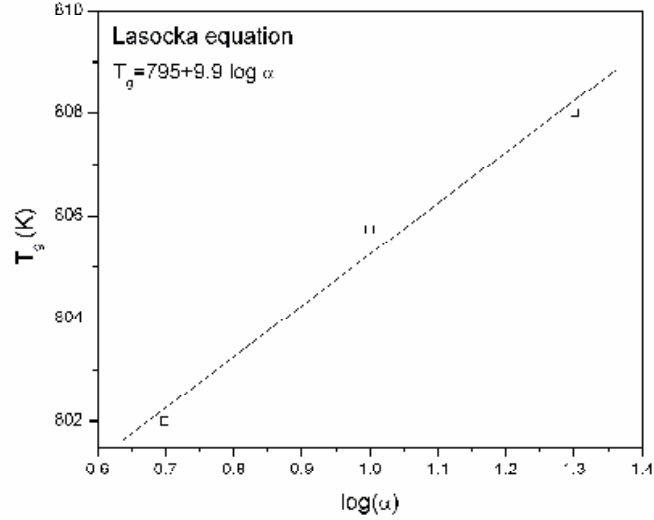

Fig. 4: $T_g$ vs. log (a) for the as-quenched BNBO glass-plates.

where $E_g$ is an activation energy for the glass transition which involved the molecular motion and rearrangement of atoms around the glass transition temperature and $R$ is the universal gas constant.

A plot of ln $(a/T_{gp}^2)$ versus $(1000/T_{gp})$ gives a linear relation, which is depicted in Fig. 5. The value of the activation energy $E_g$ obtained from the above plot is 656 ± 5 kJ/mol.

The activation energy for the glass transition $E_g$ could be calculated using the following formula, suggested by Moynihan

$$\ln\alpha = const. - \frac{E_g}{RT_{gp}} \quad (4)$$

For Eq. 4, the necessary constraint is that prior to reheating, the glass must be cooled from above to well below the glass transition region at a rate, which is either equal or proportional to the reheating rate.

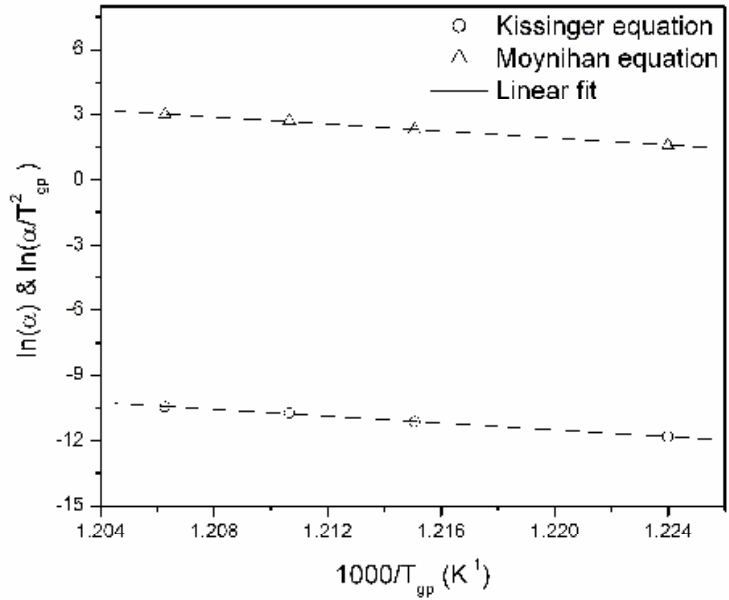

Fig. 5: ln (*a*) and ln (*a/T$_{gp}^2$*) versus 1000/*T$_{gp}$* for BNBO glass-plates.

In the present experiments, sample is heated and cooled around the glass transition region in such a way that the cooling rate is equal to the reheating rate. A plot of ln *(a)* versus (*1000/T$_{gp}$*) is shown in Fig. 5. The value obtained for $E_g$ from Fig. 5 is 648 ± 5 kJ/mol, which is close to that obtained by Kissinger's method.

**2. Glass transition and crystallization correlation**

Non-isothermal methods are useful in obtaining kinetic parameters associated with the crystallization of the glasses. This would be obtained by monitoring the shift in the position of an exotherm as a function of the heating rate. The non-isothermal crystallization process of BNBO glass powdered sample at different heating rates (5, 10, 15 and 20 K/min) exhibits a single symmetric exothermic peak in the DSC studies, as shown in Fig. 6. Similar trends were observed for BNBO glass plates of different thickness under study, which are not shown in Fig. 6.

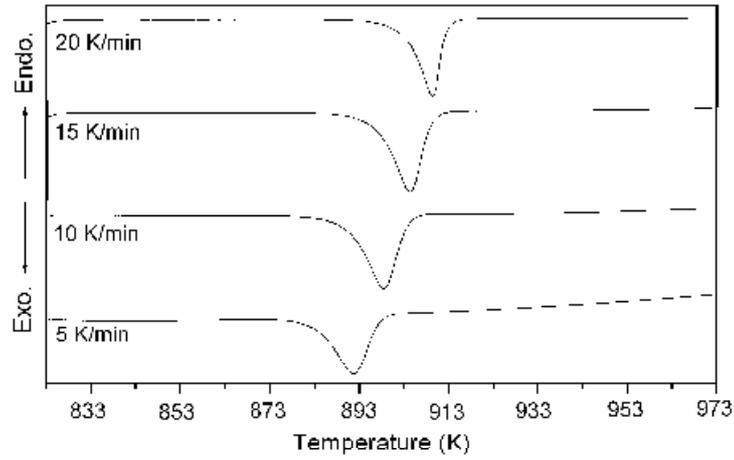

Fig. 6: DSC traces for powdered samples at various heating rates.

We notice a systematic shift in the crystallization temperature to higher temperatures with an increase in the heating rate. For rationalizing heating rate dependent crystallization temperature, Lasocka equation can be invoked, which is as follows

$$T_{cr} = A_{cr} + B_{cr} \log \alpha \qquad (5)$$

where $A_{cr}$ is the crystallization temperature at the heating rate of 1 K/min and $B_{cr}$ is a constant. Fig. 7 shows the plots of $T_{cr}$ versus log a for both the samples (**plate** and powdered). The experimental points for the present work along with the theoretical fit (solid line) to the above relation suggest its validity.

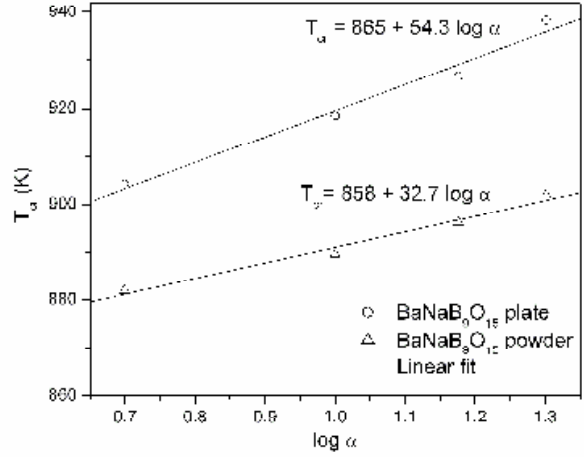

Fig. 7: $T_{cr}$ versus log $a$ plots for the as-quenched powdered and plate shaped BNBO glasses.

The above relation for BNBO plate shaped sample can be written as

$$T_{cr} = 865 + 54.3 \log \alpha \qquad (6)$$

It is to be noted that the above relation is valid for the BNBO glass plates of various thicknesses, since $T_{cr}$ is invariant with the thickness of the samples. While for the powdered sample

$$T_{cr} = 858 + 32.7 \log \alpha \qquad (7)$$

To correlate the glass transition and crystallization temperatures of BNBO glasses, one could derive the relation (using Eq. 1 and Eq. 5) in the form

$$T_g = \left( \frac{B_{cr} A_g - A_{cr} B_g}{B_{cr}} \right) + \left( \frac{B_g}{B_{cr}} \right) T_{cr} \qquad (8)$$

Therefore, for BNBO glass-plates, the above relation can be expressed as

$$T_g = 636 + 0.18 T_{cr} \qquad (9)$$

while for BNBO glass powder, one would write

$$T_g = 538 + 0.3 T_{cr} \qquad (10)$$

Eqs. (9) and (10) show that the glass transition and the crystallization temperatures are dependent parameters. Indeed similar correlation was reported for amorphous alloys by Yao *et.al.* [30]. The Lasocka equations for the glass transition and the crystallization temperatures could be used to evaluate Kauzmann temperature. The Kauzmann temperature ($T_k$) is an important parameter to characterize glassy materials from thermodynamic view point. $T_k$ is the temperature at which the entropy of liquid becomes equal to that of crystal. It denotes a lower boundary for the glass transition temperature from thermodynamic aspects and existence of an under-cooled liquid below this temperature would violate thermodynamic law [29]. Eq. 8 could be to determine the Kauzmann temperature ($T_k$) for the present glass system. The Kauzmann temperature is the lowest theoretical boundary for the glass transformation. One can assume that $T_g=T_{cr}=T_k$ [31] at a heating rate $a_k$.

After solving Eq. 8 for $T_k$ (which is equal to $T_g$ or $T_{cr}$) and $a_k$, one would arrive at

$$T_k = \frac{B_{cr} A_g - A_{cr} B_g}{B_{cr} - B_g} \tag{11}$$

and

$$\alpha_k = 10^{[(A_g - A_{cr})/(B_{cr} - B_g)]} \tag{12}$$

The values obtained for $T_k$ from Eq. 11 are around 776 K and 768 K for BNBO plate and powdered samples, respectively and the corresponding heating rates (from Eq. 12) to reach these temperatures are 0.027 K/min and 0.002 K/min. Due to very low heating rates, it is extremely difficult to observe this ideal glass transition experimentally.

## 3. Crystallization Kinetics

The isothermal crystallization process of the glass is described by the JMA equation [22-24]

$$x = 1 - \exp[-(kt)^n] \tag{13}$$

where $k$ is a rate constant which is a function of the temperature and depends on the nucleation rate and on the speed of growth of the crystallites, $n$ is the Avrami exponent which reflects the characteristics of nucleation and growth process, $t$ is the transformation time and $x$ is the volume fraction that is crystallized at time $t$. The constant $k$ is related to the crystallization activation energy, $E_c$ for the process, and its Arrhenius temperature dependence is given by

$$k = k_o \exp\left(-E_c/RT\right) \tag{14}$$

where $k_o$ is the frequency factor, $R$ is the universal gas constant and $T$ is the absolute temperature. There are various methods that are derived based on the JMA equation for non-isothermal crystallization process [32]. All the methods assume a constant heating rate, $a$, in the DSC or DTA experiments, wherein

$$T = T_i + at \tag{15}$$

where $T_i$ is the temperature at which the crystallization begins and $T$ is the temperature after time $t$. For non-isothermal crystallization, JMA equation can be written as

$$x = 1 - \exp\left[-\left(k_o \exp\left(-E_c/RT\right)\right)^n \times \left((T-T_i)/\alpha\right)^n\right] \tag{16}$$

on rearranging and taking logarithm,

$$-\ln(1-x) = \left[\left(k_o \exp\left(-E_c/RT\right)\right)^n \times \left(\frac{T-T_i}{\alpha}\right)^n\right] \tag{17}$$

for $-\ln(1-x) = 1$ or $x = 0.63$, Eq. 17 can be written as

$$k_o \left[\exp\left(-\frac{E_c}{RT_{0.63}}\right)\right] \times \left(\frac{T_{0.63} - T_i}{\alpha}\right) = 1 \tag{18}$$

where $T_{0.63}$ is the temperature at which 63% glass is crystallized.

When one expresses the above equation in logarithmic form, one obtains,

$$\ln\left(\frac{\alpha}{T_{0.63} - T_i}\right) = \ln k_o - \frac{E_c}{RT_{0.63}} \tag{19}$$

Thus the plot of ln $[a/(T_{0.63}-T_i)]$ versus $1/T_{0.63}$ yields a straight line. The values of crystallization activation energy and frequency constant are obtained from the slope and intercept of the above plot, respectively. This method is approximation-free and is applied to extract the crystallization kinetic parameters for the present glasses.

Eq. 17 could be used for calculating the Avrami exponent. For fractions of crystallization $x_1$ and $x_2$ at different temperatures $T_1$ and $T_2$ at a heating rate of $a$, one obtains

$$\frac{\ln(1-x_1)}{\ln(1-x_2)} = \left[\frac{\exp(-E_c/RT_1)}{\exp(-E_c/RT_2)} \times \frac{(T_1-T_i)}{(T_2-T_i)}\right]^n \tag{20}$$

on taking the logarithm of the above equation,

$$\ln\left[\frac{\ln(1-x_1)}{\ln(1-x_2)}\right] = n\left[\frac{E_c}{R}\left(\frac{1}{T_2} - \frac{1}{T_1}\right) + \ln\frac{(T_1-T_i)}{(T_2-T_i)}\right] \tag{21}$$

The Avrami exponent ($n$) could be obtained using the above relation. The crystallized fraction $x$ at any temperature $T$ is $x=A_T/A$, where $A$ is the total area of the exotherm, $A_T$ is the area under exotherm from $T_i$ to temperature $T$ as shown in Fig. 8. The crystallized fraction x as a function of temperature at all the heating rates under study for BNBO glass-plates (0.5 mm) and powdered samples, are shown in Figs. 9 and 10.

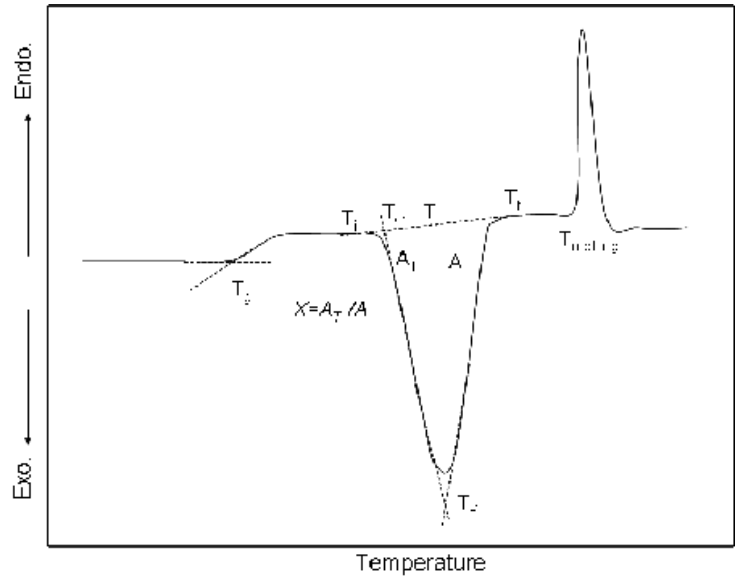

Fig. 8: Typical non-isothermal DSC plot.

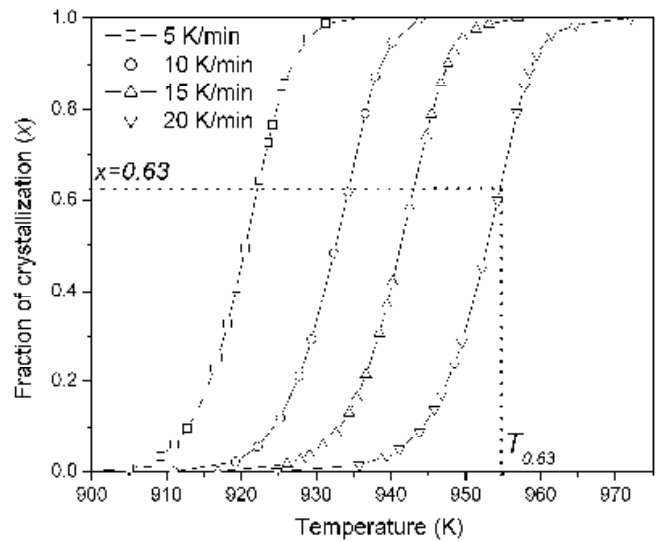

Fig. 9: Fraction of crystallization vs. temperature curves at various heating rates
for 0.5 mm thick BNBO glass-plates.

The crystallization activation energies, $E_c$ and the frequency constants for BNBO glass-plates and powdered samples, could be calculated using Eq. 19. A plot of $\ln[a/(T_{0.63} - T_i)]$ versus $1000/T_{0.63}$ for BNBO glass-plates (0.5 mm) and powdered samples are shown in Figs. 11 and 12, respectively. The values of $E_c$ obtained from the above plot are 279 ± 10 kJ/mol and 424 ± 10 kJ/mol for glass- plates and powdered samples, respectively.

Table.1: Crystallization parameters for the BNBO samples

| BNBO samples | $T_g$ ± 2K (10 K/min) | $T_{cr}$ ± 2K (10 K/min) | $?T_{FWHM}$ ± 0.5 | $E_c$ ± 10 (kJ/mol) | $n ± 0.2$ |
|---|---|---|---|---|---|
| Powdered | 806 | 890 | 5.6 | 424 | 2 |
| 0.5 mm thick plate | 806 | 918 | 10.3 | 279 | 2.5 |
| 1.5 mm thick plate | 806 | 918 | 7.2 | 279 | 3.5 |
| 2.5 mm thick plate | 806 | 918 | 6.5 | 279 | 3.9 |

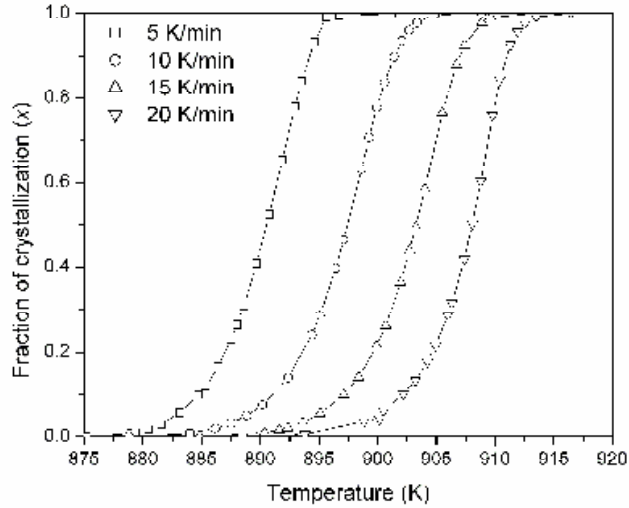

Fig. 10: Fraction of crystallization vs. temperature at various heating rates for BNBO glass powder.

It is seen that BNBO glass powder has higher activation energy for the crystallization than that of the plate shape samples. Since the specific area is higher in the case of powdered samples, more energy is required to relieve the surface strain and hence large activation energy for crystallization. The values for the frequency constant are about $2 \times 10^{15}$ and $2 \times 10^{24}$ **sec$^{-1}$** for BNBO plate and powdered samples, respectively. A large difference between the values of frequency constant for the plate and the powdered samples is due to a large difference in the activation energy [33]. It is worth mentioning that the frequency constant indicates the number of attempts per second made by the nuclei to overcome the energy barrier. The higher value of frequency constant for the powdered sample implies that these samples have greater tendency for crystallization. Chen [25] developed a method, which could be used commonly for analyzing the crystallization data in non-isothermal DSC or DTA experiments. According to which

$$\ln\left(\frac{\alpha}{T_{cr}^{2}}\right) = -\frac{E_c}{RT_{cr}} + const. \qquad (22)$$

The plots of $\ln(\alpha/T_{cr}^2)$ versus $1000/T_{cr}$ for both the samples (BNBO glass-plates and powdered) are shown in Figs. 11 and 12, respectively.

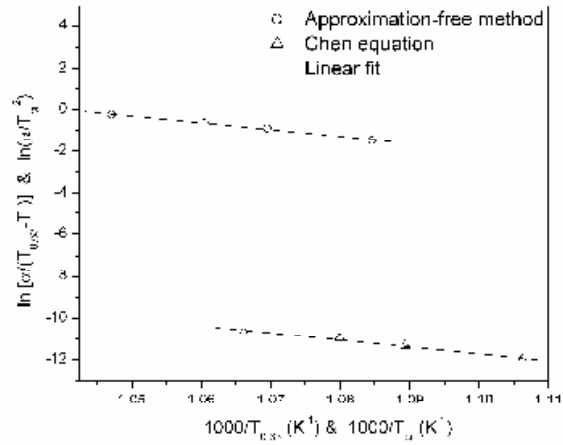

Fig. 11: $\ln [\alpha/(T_{0.63} - T_i)]$ vs $1000/T_{0.63}$ and $\ln (\alpha/T_{cr}^2)$ vs. $1000/T_{cr}$ plots for BNBO glass-plates (0.5 mm).

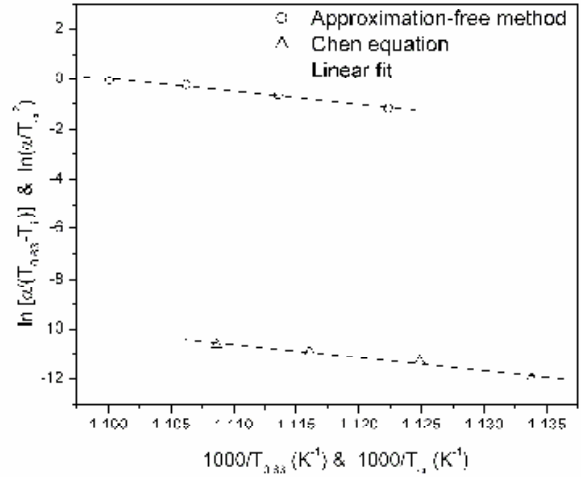

Fig. 12: ln [$a/(T_{0.63} - T_i)$] vs $1000/T_{0.63}$ and ln ($a/T_{cr}^2$) vs. $1000/T_{cr}$ plots for BNBO glass powder.

The values that are obtained for $E_c$ are 283 ± 10 kJ/mol and 432 ± 10 kJ/mol for the **plate** and powdered samples, respectively. These values are in good agreement with that obtained by the approximation-free method (Eq. 19). It is clear from the Eq. 22 that the values for the activation energy would not be different for different thicknesses of samples as $T_{cr}$ is found to be independent of the thickness of the samples (Fig. 3). The Avrami exponents for BNBO plate (0.5 mm thick) and powdered samples were determined using the Eq. 21. The values for the Avrami exponent (*n*) are 2.5 ± 0.1 and 2 ± 0.1 for the 0.5 mm thick plate and powdered BNBO glass samples, respectively. Augis and Bennett [34] pointed out that the crystallization with the same $E_c$ but different values of *n* would exhibit crystallization peaks at the same temperature. The value of *n* determines the shape of the crystallization peak; the higher the value of *n*, the narrower the peak. According to Eq. 23, a sharp peak (small ?$T_{FWHM}$, large *n*) implies bulk crystallization while a broad peak (large ?$T_{FWHM}$, small *n*) signifies surface crystallization. The following expression could be used for estimating the value for *n*:

$$n = \frac{2.5 * T_P^2}{\Delta T_{FWHM} \; E_C/R} \qquad (23)$$

where $\Delta T_{FWHM}$ is the width of the crystallization peak at half-maximum. In the present study, peak crystallization temperature ($T_P$) and activation energy associated with crystallization, $E_c$, were insensitive to the thickness of the BNBO glass plates. For the glass plates of different thicknesses, the above relation (Eq. 23) can be written as;

$n.\Delta T_{FWHM}$ = constant  (24)

It suggests that the product of $n$ and $\Delta T_{FWHM}$ is constant for glass plates of different thicknesses and can be stated as;

$(n.\Delta T_{FWHM})_{t\,1} = (n.\Delta T_{FWHM})_{t\,2}$  (25)

We have calculated the values of $n$ for 1.5 and 2.5 mm thick BNBO glass plates using the values of $\Delta T_{FWHM}$ for all the glass plates along with the value of $n$ obtained for the 0.5 thick glass plates. The obtained values for $n$ are reported in the table. 1. The value of $n$ generally indicates the mode of crystallization. It is suggested that $n=4, 3$ and $2$ correspond to the volume nucleation, three, two and one-dimensional growth, respectively and $n=1$ suggests surface nucleation, and one-dimensional growth. In the present study, the $n$ values for the glass-plates indicate volume nucleation accompanied by one, two and three-dimensional growth depending on the thickness of the samples, while for the powdered sample, one-dimensional bulk growth is dominated.

IV. Conclusions

The glass transition and crystallization behaviour are rationalized in terms of Lasocka formulae. The linear correlations are found between the glass transition and crystallization temperatures and are given as $T_g = 636 + 0.18\ T_{cr}$ and $T_g = 538 + 0.3\ T_{cr}$ for BNBO glass plates and powdered samples, respectively. The values of Kauzmann temperature are calculated using above relations representing the lower bound for the kinetically observed glass transition. The crystallization parameters for BNBO glasses, which are scientifically important, have been analyzed using new approximation free method derived for non-isothermal experiments. The average value for the crystallization activation energy is $281 \pm 10$ kJ/mol for BNBO glass plates and is $424 \pm 10$ kJ/mol for powdered samples. Crystallization temperature and activation energy were found to be independent of the thickness of the BNBO glass plates. However, the value obtained for the Avrami exponent increases with increase in thickness, indicating that the thickness of the samples has a strong influence on the crystallization mechanisms. These thermal parameters are crucial for the fabrication of glass nano/microcrystal composites.

**References**

1. G. Senthil Murgan and K.B.R. Varma 2002 *J. Mater. Chem.* **12** 1426.

2. M. M. Layton and J. W. Smith 1975 *J. Am. Ceram. Soc.* **58** 435.

3. J. Zhang, B. I. Lee, and Robert W. Schwartz 1999 *J. Appl. Phys.* **85** 8343.

4. N. Syam Prasad, K. B. R. Varma, Y. Takahashi, Y. Benino, T. Fujiwara and T. Komatsu 2003 *J. Solid State Chem.* **173** 209.

5. G. Senthil Murugan and K. B. R. Varma 1999 *Mater. Res. Bull.* **34** 2201.

6. T. Komatsu, R. Ihara, T. Honma, Y. Benino, R. Sato, H.G. Kim and T. Fujiwara 2007 *J. Am. Ceram. Soc.* **90** 699.

7. P. Gupta, H. Jain, D. B. Williams, T. Honma, Y. Benino, and T. Komatsu 2008 *J. Am. Ceram. Soc.* **91** 110.

8. N. S. Prasad and K. B. R. Varma 2005 *J. Am. Ceram. Soc.* **88** 357.

9. T. Honma, Y. Benino, T. Fujiwara, T. Komatsu and R. Sato 2005 *J. Am. Ceram. Soc.* **88** 989.

10. T. Sasaki, Y. Mori, M. Yishimura, Y. K. Yap, and T. Kamimura 2000 *Mater. Sci. Eng.* **30** 1.

11. T. Sugawara, R. Komatsu and S. Uda 1998 *Solid State Commun.* **107** 233.

12. Y. Wu, J. Liu, P. Fu, J. Wang, H. Zhou, G. Wang and C. Chen 2001 *Chem. Mater.* **13** 753.

13. Jun-Ming Tu and D. A. Keszler 1995 *Mater. Res. Bull.* **30** 209.

14. N.Penin, L. Seguin, M. Touboul, and G. Nowogrocki 2001 *Int. J. Inorg. Mater.* **3** 1015.

15. R. Vaish and K.B.R. Varma 2008 *J. Am. Ceram. Soc.* **91** 1952.

16. A. Goel, E. R Shaaban, M. J Ribeiro, F C L Melo and José M F Ferreira 2007 *J. Phys.: Condens. Matter.* **19** 386231.

17. M. Saxena 2005 *J. Phys. D: Appl. Phys.* **38** 460.

18. S. Venkataraman, H. Hermann, C. Mickel, L. Schultz, D.J. Sordelet and J. Eckert 2007 *Phys. Rev.B* **75** 104206.

19. A.Goel, E.R. Shaaban, F.C.L. Melo, M.J. Ribeiro and J.M.F. Ferreira 2007 *J. Non-Cryst. Solids* **353** 2383.

20. A. Arora, A. Goel, E.R. Shaaban, K.Singh, O.P. Pandey and J.M.F. Ferreira 2008 *Physica B*, **403** 1738.

21. D.M. Minic and B. Adnadevic 2008 *Thermochimica Acta,* **474** 41.

22. M. Avrami 1939 *J. Chem. Phys.* **7** 1103.

23. M. Avrami 1940 *J. Chem. Phys.* **8** 212.

24. M. Avrami 1941 *J. Chem. Phys.* **9** 177.

25. H. S. Chen 1978 *J. Non-Cryst. Solids* **27** 257.

26. H. E. Kissinger 1956 *J. Res. Nat. Bur. Stand.* **57** 217.

27. C. T. Moynihan, A. J. Easteal, and J. Wilder 1974 *J. Phys. Chem.* **78** 2673.

28. M. Lasocka 1976 *Mater. Sci. Eng.* **23** 173.

29. W. Kauzmann 1948 *Chem. Rev.* **43** 219.

30. B. Yao, H. Ma, H. Tan Y. Zhang and Y. Li 2003 *J. Phys.: Condens Matter* **15** 7617.

31. T. Ichitsubo, E. Matsubara, H. Numakura, K. Tanaka, N. Nishiyama and R. Tarumi 2005 *Phys. Rev. B* **72** 052201.

32. H. Yinnon and D.R. Uhlmann 1983 *J. Non-Cryst. Solids* **54** 253.

33. N. Koga and J. Šesták 1991 *Thermochimica Acta,* 182 201.

34. J.A. Augis and J.E. Bennett 1978 *J. Therm. Anal.* 13 283.

**Figure Captions**

Fig.1: XRD pattern for the as-quenched pulverized BNBO glass.

Fig. 2: Fig. 2: DSC traces for the as-quenched samples of BNBO glass powder and plates of different thicknesses.

Fig. 3: Glass transition peaks for BNBO glass-plates at various heating rates.

Fig. 4: $T_g$ vs. log ($a$) for as-quenched BNBO glass-plates.

Fig. 5: ln ($a$) and ln ($a/T_{gp}^2$) versus 1000/$T_{gp}$ for BNBO glasses.

Fig. 6: DSC traces for powdered samples at various heating rates.

Fig. 7: $T_{cr}$ versus log $a$ plots for the as-quenched powdered and plate shaped BNBO glasses.

Fig. 8: Typical non-isothermal DSC plot.

Fig. 9: Fraction of crystallization vs. temperature curves at various heating **rates for 0.5 mm thick** BNBO glass-plates .

.Fig. 10: Fraction of crystallization vs. temperature at various heating rates for BNBO glass powder.

Fig. 11: ln [$a/(T_{0.63} - T_i)$] vs 1000/$T_{0.63}$ and ln ($a/T_{cr}^2$) vs. 1000/$T_{cr}$ plots for BNBO glass-plates (0.5 mm).

Fig. 12: ln [$a/(T_{0.63} - T_i)$] vs 1000/$T_{0.63}$ and ln ($a/T_{cr}^2$) vs. 1000/$T_{cr}$ plots for BNBO glass powder.